# A high-resolution gridded inventory of coal mine methane emissions for India and Australia

Pankaj Sadavarte[1,2,*], Sudhanshu Pandey[1], Joannes D. Maasakkers[1], Hugo Denier van der Gon[2], Sander Houweling[1,3], Ilse Aben[1]

[1]SRON Netherlands Institute for Space Research, Leiden, The Netherlands.

[2]Department of Climate, Air and Sustainability, TNO, Utrecht, The Netherlands.

[3]Department of Earth Sciences, Vrije Universiteit, Amsterdam, The Netherlands.

*p.sadavarte@sron.nl

## Abstract

Coal mines are globally an important source of methane and also one of the largest point sources of methane. We present a high resolution 0.1° × 0.1° bottom-up gridded emission inventory for methane emissions from coal mines in India and Australia, which are among the top five coal producing countries in 2018. The aim is to reduce the uncertainty in local coal mine methane emissions and to improve the spatial localization to support monitoring and mitigation of these emissions. For India, we improve the spatial allocation of the emissions by identifying the exact location of surface and underground coal mines and we use a tier-2 Intergovernmental Panel on Climate Change (IPCC) methodology to estimate the emissions from each coal mine using country-specific emission factors. For Australia, we estimate the emission for each coal mine by distributing the state-level reported total emissions using proxies of coal production and the coal basin-specific gas content profile of underground mines. Comparison of our total coal mine methane emission from India with existing global inventories showed our estimates are about a factor 3 lower, but well within range of the national Indian estimate reported to United nations framework convention on climate change (UNFCCC). For both the countries, the new spatial distribution of the emissions show large difference from the global inventories. Our improved emissions dataset will be useful for air quality or climate modelling and while assessing the satellite methane observations.

# 1 Introduction

Methane ($CH_4$) is the second most important greenhouse gas (GHG), after carbon dioxide ($CO_2$). Due to its shorter lifetime of about ~10 years, mitigation of $CH_4$ along with other short-lived climate pollutants (SLCPs) is of utmost importance to tackle the global warming on a relatively shorter time scale (Shindell et al., 2012). Methane emissions from coal mines are a significant source, estimated to be 43 (31-62) Tg, which is 11% of annual total anthropogenic emissions (Saunois et al., 2020). The range indicates a global uncertainty of -30% - +45%. However, the uncertainty of individual coal mine emissions is much larger than in the global budget. This has been highlighted further by recent studies using satellite observations to asses coal mine emissions (Varon et al., 2020, Sadavarte et al., 2021).

Recently, satellites like TROPOMI (TROPOspheric Monitoring Instrument) and GHGSat have been shown to be able to successfully detect and quantify methane emissions from large point sources from the oil and gas sector (de Gouw and Veefkind, 2020, Pandey et al., 2019, Varon et al., 2019, Cusworth et al., 2021, Irakulis-Loitxate et al., 2021) as well as from coal mines (Varon et al., 2020, Sadavarte et al., 2021). Varon et al. (2020) for the first time showed methane emissions from individual coal mine vents using space-based observations from GHGSat-D satellite, with estimated emission fluxes varying from 2.4-5.9 ton h-1. Sadavarte et al. (2021) detected three methane hot spots over the Bowen basin in Australia using TROPOMI observations, showing coal mine methane emissions ranging from 17-26 ton h-1 per source location. These initial studies show the potential of satellite observations to significantly improve our knowledge on large methane point sources, such as coal mines in the near future. The rapid developments in satellite remote sensing of methane also emphasize the need for high resolution spatially resolved a priori information on the location and bottom-up estimated emissions of coal mine methane (CMM) sources, to guide and support their data analysis. This is a relatively new development for (national) GHG inventories because detailed spatio-temporal inventories previously had limited added value due to the rapid mixing and long lifetime of GHGs. This is very different from air pollutant inventories where this is needed to, for example, model population exposure.

Globally, China is the largest producer of coal followed by India, US and Australia (Global Energy Statistical Yearbook 2019) (Figure S1). The countries under Annex-I estimate and report methane emissions from coal mining to the UNFCCC, i.e., Australia and the US (UNFCCC, 2019). Next to this official UNFCCC report, various studies have produced (partly) independent estimates of methane emissions from coal mining (Table S1) (Janssens-Maenhout et al., 2019, Höglund-Isaksson, 2012, GMI, 2010, Scarpelli et al., 2020, Singh et al., 2015, Sheng et al., 2019, Hoesly et al., 2018). As indicated by the emission estimates for the top-5 coal producing countries (Table S1), there is a considerable difference among the methane emission from global coal mining activities. Hence reduction of that uncertainty would be highly beneficial for better constrained emission estimates derived from the emission flux inversions using atmospheric measurements from satellite observations and detailed high-resolution bottom-up inventories.

Our overarching ambition is to improve both the spatial localization of coal mine methane emissions that allows top-down analyses as well as reduce the uncertainty in local to national CMM emissions by using TROPOMI data. A first pre-requisite is having good prior CMM emission inventory at high spatial resolution to guide the satellite data analysis. Among the top-5 coal producing countries, gridded high resolution CMM emissions were produced for China and the USA by Sheng et al., (2019) and Maasakkers et al., (2016), which are respectively the first and third largest coal producing countries (Figure S1). These regional studies have shown significant improvements in national estimates and spatial distribution compared to the more generic global emission inventories. For example, using a regional database and emission factors, Sheng et al., (2019) showed that CMM emissions for China are overestimated in EDGARv4.3.2 by a factor 1.2, and the number of coal mines in EDGARv4.3.2 used for allocation was just ~45% of the actual coal mines present updating which improved the overall spatially gridded CMM emissions over China. Therefore, as a first step towards our goal to improve the global CMM emissions data we review and develop a spatially resolved high-resolution emission inventory for two other high ranked coal producing countries, India and Australia for 2018. In this bottom-up study, emissions are estimated from coal production and post-mining handling activities of surface and underground mines. This study does not account for any pyrolysis-based methane emission from burning of fossil fuels used during mining activities. Section 2 explains the coal production activities for India and Australia, section 3 describes the emission factors used in this study, section 4 describes our spatial distribution method and section 5 provides the final estimates.

## 2 Emission estimation method and activity rate

We estimate the coal mine methane emission as the product of annual coal produced and the emission factor (Volume 2 Chapter 4 IPCC, 2006a). For India, we calculate the emissions using below equation (1).

$$E_{i,j} = C_{i,k} \times EF_{i,j,k} \times (1 - R) \qquad (1)$$

Where $E$ (g) is the methane emissions, $C$ (kg) is the amount of coal produced by each mine, $EF$ is the emission factor in g of $CH_4$ per kg coal produced. $EF$ is a function of the type of mine $i$ (surface or underground), emission activity $j$ (mining or post-mining), and type of coal $k$ (hard or brown coal). With reference to coal mine methane, brown coal has a lower methane content compared to hard coal due to the lower rank of coal and lower carbon content. $R$ is the methane removal efficiency (%) due to abatement technologies such as flaring, on-site oxidation or capture and transfer.

For Australia, the methane emissions data are available at state-level under categories of surface and underground mines from national inventory report 2018 (NIR 2018, 2020). For underground mines, we distribute methane emissions to individual coal mines using the raw coal production from each mine and the gas content profile of the basin (equation 2a-c). For surface mines, no additional information such as gas content is available and hence the emissions are distributed based on coal production data by mine only as shown in equation 2d.

$$frac_{basin} = \frac{G_{basin} \times C_{basin}}{\sum_a G \times C} \qquad (2a)$$

$$E_{basin} = E_{state,ug} \times frac_{basin} \qquad (2b)$$

$$E_x = E_{basin} \times \frac{c_{x,ug}}{C_{basin}} \qquad (2c)$$

$$E_y = E_{state,oc} \times \frac{c_{x,oc}}{C_{state,oc}} \qquad (2d)$$

Where $G_{basin}$ is the gas content of the coal basin provided as ton $CO_2$-eq per ton of coal in national inventory report (NIR 2018, 2020), $C_{basin}$ is the total coal produced by underground mines in that basin. $\sum_a G \times C$ is the total product of gas content and coal produced by each basin, to calculate the fraction. $frac_{basin}$ is the fraction calculated to estimate the emissions from a specific basin $E_{basin}$ using state-level underground coal mine emission $E_{state,i}$. This emission $E_{basin}$, is further split to individual underground (ug) mines using the ratio of coal produced in each mine $x$ to $C_{basin}$. For surface mines, the individual mine emission $E_y$ is calculated using state-level emissions from surface mines ($E_{state,oc}$) and the ratio of coal produced in each mine $x$ to $C_{state,oc}$, where, $C_{state,oc}$ is the total coal produced in surface mines.

## 2.1 Coal production in India

India produced 773 million tons of raw coal during 2018, comprised of 94% hard coal (bituminous/sub-bituminous) and 6% brown coal (lignite) (MoC, 2019). The brown coal mining occurs in the states of Gujarat, Tamilnadu and Rajasthan while hard coal mining is found in Assam, Andhra Pradesh, Maharashtra, Madhya Pradesh, Jharkhand, Chhattisgarh, West Bengal, Telangana, Orissa, and Meghalaya (MoC, 2019) (Figure 1). The northeastern state of Meghalaya is known for its so-called "rat-mining" where small holes are dug and the coal is excavated using man power, while mechanized surface mining and underground mining takes place elsewhere in India. In 2018, 94% of annual coal was produced from surface mines and only 6% from underground mines. This has changed over the last two decades, i.e. earlier in 1998, 77% of the annual coal production came from surface mines and 23% from underground mines (Figure S2). About 225 surface, 205 underground and 24 mixed mines are reported to be operating in India during 2018-2019 (MoC, 2018). A list of operating underground mine was available for 2014-15 along with its coal production data (LS, 2020). We assume these underground mines to be active and operating in 2018-19. For this study, we scale the coal production in each of the underground mines proportionally using national underground coal produced in 2018-2019 (MoC, 2019). In case of surface mines, the annual coal produced for surface mine was available at state-level for 2018-19 (MoC, 2019), while mine-level details were also compiled from publicly available Global Coal Mine Tracker database (GCMT, 2021).

## *2.2 Coal production in Australia*

In 2018, Australia produced 618 million tons of raw coal, out of which 91% is hard coal and the rest is brown coal. In terms of mine type, 77% of the coal is produced in surface mines and rest in underground mines. The Australian coal production has seen an increase of 70% over the last two decades since 1998 (Figure S3). About 91% of the national coal production occurs in Queensland and New South Wales states while the remaining 9% comes from Western Australia, Victoria and Tasmania (Figure 2). In 2018, there were 116 operating coal mines out of which 43 were underground and 73 surface, located across Australia (NIR 2018, 2020). Until 2018, a total 119 mines have been decommissioned and the fugitive emissions from these mines are accounted for as well (NIR 2018, 2020). The coal mine details including operating mines, mine type, coal production at state-level and coal type were compiled from the Australian mine atlas and Australian energy statistics (Australian Government, 2020, DoEE, 2019). Besides the significant number of decommissioned mines, there are mines that have undergone rehabilitation for care and maintenance and no longer produce coal. Hence, each mine/company's annual review report and its mining operation plans were scrutinized carefully for raw coal production data and its current operating status especially for New South Wales, Victoria, Tasmania and Western Australia states. For Queensland state, the coal production details were compiled from the official website (Open data portal, 2020).

# 3 Coal mine methane emission factors

Methane trapped in coal can be extracted and used for energy production or it can be vented into the atmosphere or flared for safety reasons (Moreby et al., 2010). Prior to the underground mining, the methane extracted from the coal seam is termed as coal seam gas (CSG) or coal bed methane (CBM). Broadly, the methane released during the surface and underground mining operations is known as coal mine methane (CMM). Methane vented from the underground mines through ventilation shafts for safety and better air circulation is known as ventilation air methane (VAM). The following section describes the regional emission factors used in this study for methane emissions.

## *3.1 India*

The region-specific emission factors, i.e., tier-2 method of the Intergovernmental panel on climate change (IPCC) are used here to estimate methane from coal mines. The Central Institute of Mining and Fuel Research (CIMFR) measured emission factors for 16 surface and 83 underground mines. Multiple studies and national reports have used these emission factors for estimating CMM emissions (MoEFCC, 2015, 2018, 2021, Singh et al., 2015). The emission factors for underground mines are further classified based on the degree of gassiness (Singh et al., 2015). In India, the underground mines are classified as degree-I/II/III gassy mines depending on the methane emission rate of < 1 m3 ton-1, 1-10 m3 ton-1 and > 10 m3 ton-1 coal (MoLE, 2017). The emission factors for surface mines are measured as the methane flux (m3 hr-1) collected in a rectangular chamber from the coal surface exposed over a period of time. The methane content in the chamber is determined using gas chromatography (MoEFCC, 2015, Singh et al., 2015). For underground mines, the emission factor (m3 min-1) is derived using the methane content of the air sample collected in the ventilation shaft, by knowing the cross-sectional area of the shaft and the velocity of the air passing through the shaft. The methane flux from surface and underground mine was later divided by the daily coal produced to convert the units to m3 ton-1 coal produced (MoEFCC, 2015, Singh et al., 2015). The emission factor for brown coal was considered to be one-fourth of black coal similar to Kholod et al., (2020). According to MoLE (2017) during 2015, 260 underground mines were classified as degree-I gassy, 71 underground mines as degree-II gassy and 7 underground mines as degree-III gassy. Even for post-mining activities in underground mines, the emission factors are categorized as per the degrees of gassiness. For surface mines no information is available on their gassiness level and therefore the average emission factor from measurements was used throughout the surface mines for estimation (Singh et al., 2015). As there were no mitigation programs reported to capture and utilize fugitive methane emissions from coal mines during 2018, methane from all mining activities is assumed to be released into the atmosphere unabated.

### *3.2 Australia*

The methodology on fugitive emissions from solid fuel i.e. coal mining is documented in the Australian National Inventory Report to the United Nations Framework Convention on Climate Change (NIR 2018, 2020). Methane emission from coal mines are accounted for from operating mines, decommissioned mines, post mining and flaring activities. The coal fields in Australia are rich in gas content which varies from 0.4 kg $CH_4$ ton-1 in Western basin to ~12 kg $CH_4$ ton-1 in Southern basin (NIR 2018, 2020). Under National Greenhouse and Energy Reporting (NGER) guidelines, each underground mine is obliged to measure and report their annual emissions to the Australian National Greenhouse Accounts. Their method of estimation follows a IPCC tier-3 approach, which is the most detailed estimation method. For surface mines, both tier-2 and tier-3 based methodologies are used for estimating emissions. The tier-2 default emission factors used for surface mining ranges from 1 m3 ton-1 to 3.2 m3 ton-1 for black coal and 0.0162 m3 ton-1 for brown coal (NIR 2018, 2020). Tier-3 emission factors are used for each surface mine in the Gunnedah, Western, Surat, Collie, Hunter and Newcastle basin in New South Wales state. With such detailed reporting in place we rely on the state-wise coal mine emissions from National Inventory Report 2018 and use the primary activity data on coal production and gas content profile for reconstructing a bottom-up emission estimate for each individual mine. The main reason for doing this is that the individual mine emissions are not publicly available but are reported in aggregated form at state level. These volume based emission factors can be converted to mass units by multiplying them with the density of methane at normal temperature and pressure, i.e. 20°C and 1 atmospheric pressure.

# 4 Spatial proxies for emission distribution

The methane emissions were spatially allocated to the identified locations of underground and surface coal mines in India and Australia (Figure 1 and 2).

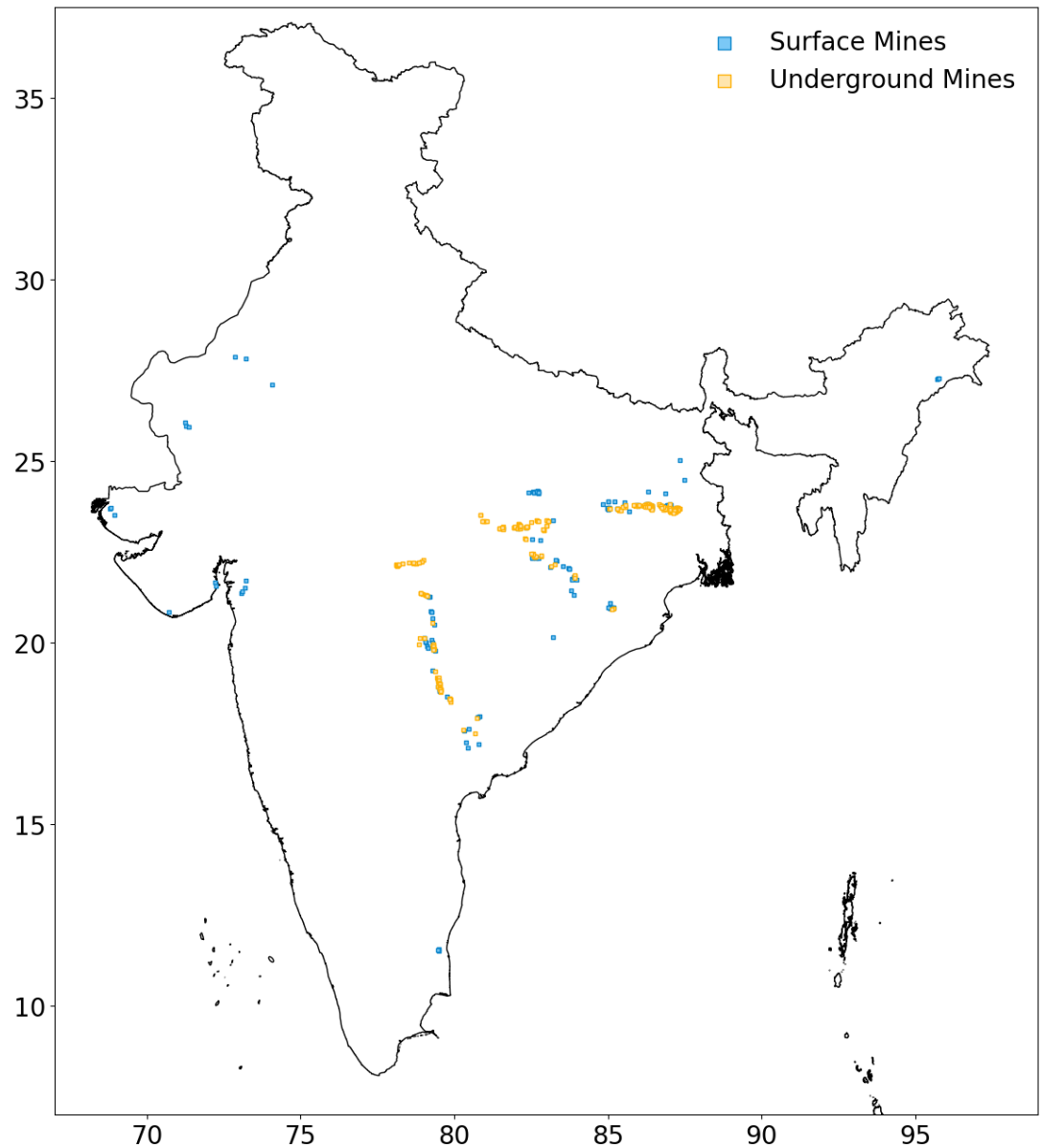

**Figure 1. Locations of underground and surface coal mines in India identified in this study.**

The blue pixels denote the location of surface mines and the yellow pixels denote the location of underground mines. These locations were identified using digital platforms like Google Earth and Sentinel-2 satellite images, and annual review reports.

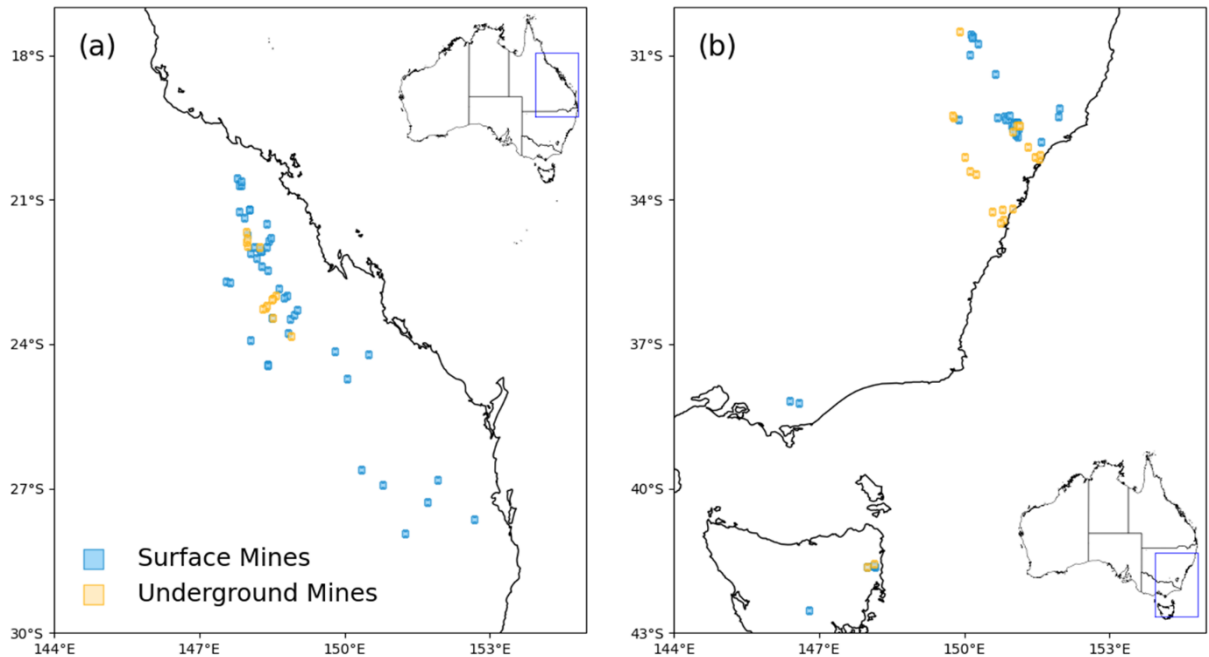

**Figure 2. Locations of underground and surface coal mines in Australia identified in this study.**

Figure (a) shows the location of operating underground and surface coal mines in Queensland state and (b) in New South Wales, Victoria and Tasmania state.

The coal mines were identified visually using satellite images from the Google Earth platform (Google Earth, 2020), Sentinel-2 images (Sentinel hub, 2020), and data portals (SSRC, 2020, AUSGIN, 2020, Gem Wiki, 2020). Relative to underground mines, a surface mine is easily identifiable with its distinguishable contours and the presence of coal in the vicinity of the mining area (Figure S4a). These locations were further confirmed using reports such as mining operation plans and environmental clearance documents that clearly depict the extent of mines on geographical maps and geological plates (CMPDI, 2018a, b). Underground mines appear different in Australia and India, depending on the mining method. In case of Australia, mostly the longwall underground mining method is used, in which the coal is extracted from rectangular blocks of certain depth, length and width (in meters). Many corresponding rectangular panels adjacent to each other are seen on the earth surface from space and the mining type can be verified with the mining operation plan and AUSGIN, (2020) geospatial mining portal (Figure S4b). In India, the board and pillar underground mining method is followed. A grid of tunnels is created from which coal is extracted from panels leaving behind pillars of coal that support the mine roof. The salient feature of an underground mine in India includes an inlet to the mine through an elevator with infrastructure supporting an outlet for lifting coal to the surface using conveyors or mine rail carrying coal tubs with tons of capacity. Even in this case, the spillage of coal giving a dark background was quite evident to identify an underground mine in India, as shown in Figure S4c. In our study we do not account for Australian abandoned mine emissions (and their spatial distribution) as they are only 0.02% of the national estimate. For spatial comparison the locations are gridded on a spatial resolution of 0.1° × 0.1°, similar to the global EDGAR emission inventory (Janssens-Maenhout et al., 2019).

# 5 Results and discussions

## 5.1 *Emission estimates and comparison*

Following the described bottom-up methodology, we estimate 825 Gg yr-1 fugitive methane emission in 2018 from the Indian coal mining sector. Emissions of similar magnitude were estimated for 2012 (770 Gg yr-1, Singh et al., 2015), 2014 (788 Gg yr-1, MoEFCC, 2018) and, 2016 (842 Gg yr-1, Scarpelli et al., 2020; 815 Gg yr-1, MoEFCC, 2021). From the 825 Gg yr-1 total emission, 76% methane is released due to surface mining (94% of Indian coal produced) and 24% from underground mining (6% of Indian coal produced). On a state level, Chhattisgarh is the largest producer of coal (21%) and also the largest coal mine methane emitter (18%) (Table 1). For other Indian states, coal production and their respective emissions are proportional, except for West Bengal, Madhya Pradesh and Telangana which have relatively higher emissions. One of the crucial differences lies in the number of high methane emitting mines, i.e. of degree-II and -III gassiness, in West Bengal and Madhya Pradesh. Secondly, the fraction of underground coal produced is relatively high in these states (0.09-0.34) compared to the rest of Indian states (0.006-0.056). Since the emission factor for degree-I underground mines is a factor 3 higher than the emission factor for surface mines, states with relatively large shares of underground mining will have relatively higher emissions.

**Table 1. Annual coal production and estimated methane emissions for India in 2018.**

|  | Coal production (million tons) |  | Methane emissions (Gg yr-1) |  |
| --- | --- | --- | --- | --- |
| **Statewise details** | Underground | Surface | Underground | Surface |
| Assam |  | 0.78 |  | 0.684 |
| Chhattisgarh | 8.53 | 153.36 | 20.1 | 133.8 |
| Jharkhand | 3.72 | 131.19 | 12.3 | 114.5 |
| Madhya Pradesh | 9.88 | 108.78 | 43.3 | 94.9 |
| Maharashtra | 2.00 | 47.82 | 4.4 | 41.7 |
| Orissa | 0.87 | 143.44 | 6.0 | 125.1 |
| Telangana | 9.18 | 55.98 | 23.4 | 48.8 |
| Uttar Pradesh |  | 20.28 |  | 17.7 |
| West Bengal | 8.37 | 24.54 | 86.2 | 21.4 |
| Tamilnadu |  | 23.04 |  | 20.10 |
| Gujarat |  | 12.57 |  | 10.96 |
| Rajasthan |  | 8.68 |  | 7.57 |
| National total | 42.54 | 730.45 | 195.81 | 637.31 |

The EDGAR emission inventory reported an annual emission of 2.4 Tg yr-1 (EDGARv4.3.2) and 2.7 Tg yr-1 (EDGARv5.0) from Indian coal mining for 2012 and 2015, respectively (Janssens-Maenhout et al., 2019) (Table S1). These are a factor 3.0 and 3.4 higher than our estimate mainly due to the use of higher IPCC default emission factors. The implied emission factor used in this study is 1.04 kton per million ton coal while EDGAR uses a value 3.9 kton per million ton coal (Table S2). The implied emission factors facilitate a comparison to other methods (Table S2). Similarly, the GAINS model estimates Indian CMM at 1.6 Tg yr-1 (Table S1) which corresponds to an implied emission factor of 3 kton per million ton coal (Table S2) (Höglund-Isaksson, 2012). The GAINS model assumed all lignite mines to be surface mines and bituminous/anthracite mines to be underground mines with sparse information available on coal production in India during 2005 (Personnel Communication with Höglund-Isaksson). This led to the assumption of 93% coal from underground mines instead of surface mines (Table 1 and section 2.1). Based on the coal mine details from coal directory of India (MoC, 2019) we conclude that in the Höglund-Isaksson (2012) paper, the GAINS model overestimates the role of underground coal mining in India. Compared to global IPCC default values and Kholod et al., (2020), the implied emission factor for underground mines from our study is a factor > 2.5 lower for mines as deep as 200-400 m. The most likely reason for this discrepancy is that Indian underground coal contains relatively little methane (low gassiness). This can also be concluded based on large number of degree-I gassy mines compared to degree-II and -III. However, the implied emission factor for surface mines is in agreement with the IPCC value i.e. 0.87 kton $CH_4$ per million ton raw coal (Table S2).

As per the Australian national inventory reporting, methane emissions from coal mining activities were 972 Gg yr-1, of which 73% are emitted by underground mines and 27% by surface mines. The fraction of coal produced by underground mining is only 17% but the corresponding emissions account for 73% of national CMM emissions (Table S2). The methane concentrations in each underground mine are measured and reported, thus considered to be a tier-3 approach as per IPCC guidelines. The state-wise coal produced and methane emissions with specified contributions from surface and underground mining is given in Table 2. It can be seen that Queensland state contributes 51% and New South Wales contributes ~49% of the national coal mine methane. The surface mines in Victoria are well known for their lignite coal production and contribute < 0.5% to total emissions. Overall, if the implied emission factors are compared for India and Australia, it can be inferred that the Australian underground mines emit 1.4 times more methane than Indian underground mines per unit of coal despite possible mitigation measures like flaring or oxidation. On the other hand, Indian surface mines have an implied emission factor which is 1.6 times higher than that of Australian surface mines. Australia controls the direct release of methane emission from underground mines by flaring, oxidizing and use of CMM in power stations. This is accounted for in the National Inventory Report (NIR 2018, 2020) and the final emission used in this study. However, this information is difficult to access and not easily available in public domain. Hence we cannot distinguish mines with more or less mitigation technologies in our spatial emission distribution methodology.

**Table 2. Annual coal production and national reported solid fuel fugitive methane emissions for Australia in 2018.**

|  | Annual coal production (million tons per year) | | | | | |
|---|---|---|---|---|---|---|
|  | Queensland | New South Wales | Victoria | Western Australia | Tasmania | National |
| **Raw Coal production** | | | | | | |
| Underground | 48.24[a] | 64.29[b] | | | | 112.53[c] |
| Surface | 269.12[a] | 183.83[b] | 45.90[d] | 6.65[d] | 0.38[d] | 505.89[c] |
| Total | 317.36[a] | 248.12[b] | 45.90[d] | 6.65[d] | 0.38[d] | 618.41[c] |
| **Saleable Coal production** | | | | | | |
| Underground | 29.28[a] | 53.55[b] | | | | |
| Surface | 221.94[a] | 139.81[b] | | | | |
| Total | 251.22[a] | 193.36[b] | | | | |
| Saleable coal production by Australia Energy Statistics | 250.57[e] | 196.61[e] | 43.32[e] | 6.27[e] | 0.36[e] | |
|  | Annual methane emissions (Gg per annum) | | | | | |
| Underground | 343.88[f] | 366.53[f] | | | | 710.4[f] |
| Surface | 196.26[f] | 61.06[f] | 0.50[g] | 4.50[g] | 0.26[g] | 262.6[h] |

[a] Department of Natural Resources and Mines, Queensland, Summary of the raw and saleable coal of individual mines by financial year, https://www.data.qld.gov.au/dataset/27fefb68-dc98-4300-85b6-465f0df233a8/resource/9c3c1aaf-0afa-4e58-b67c-75c0d3574abd/download/production-by-individual-mines.xlsx.
[b] Raw and saleable coal production data compiled from annual review report of the respective individual coal mines for the year 2018.
[c] National raw coal production for year 2018, common reporting format (CRF) as reported to UNFCCC for year 2018.
[d] Raw coal production estimated for states of Victoria, Western Australia and Tasmania, by proportionately distributing the raw coal (calculated as difference between National and, Queensland and New South Wales) as per the saleable coal for the respective three states.
[e] Department of Industry, Science, Energy and Resources, Australian Energy Statistics, Table P, September 2020.
[f] Methane emissions reported to UNFCCC for 2018 (source: https://ageis.climatechange.gov.au/)
[g] Estimated using bottom-up emission factors available for surface mines in the national inventory report 2020 (Table 3.32) for Victoria and Tasmania state. For Western Australia, the emission factor similar to Tasmania was used for estimating emissions.
[h] Sum of surface based coal mine emissions.

A simple tier-1 uncertainty on national emissions was estimated separately for surface and underground mines following the IPCC uncertainty assessment guidelines (Volume 1 Chapter 3 IPCC, 2006b). The combined uncertainty on total emissions was calculated by propagating uncertainties on coal production and emission factor in the categories of surface and underground mines and adding them in quadrature. For Australia, the 2018 national inventory report provided the uncertainty on the national estimates, calculated separately for surface mines as ±10.2%, underground mines as ±31.5% and abandoned mines as ±50.3%. The combined uncertainty on total coal mine methane emissions was calculated as ±11.2% following IPCC (2006b). For India, in the absence of the underlying uncertainties in the measured emission factors by CIMFR, we consider 100% for both surface and underground mines as per the uncertainty provided in third biennial update report (MoEFCC, 2021). A default uncertainty of 10% for coal production was considered consistent with the global inventory of EDGAR (Solazzo et al., 2021). Using the IPCC guidelines, ±100.5% uncertainty was propagated on both the categories of surface and underground mines. Finally a combined uncertainty of ±80% was calculated on the national coal mine methane emissions following the IPCC method of combining uncertainties (Volume 1 Chapter 3 IPCC 2006b). Table S3 provides the underlying uncertainty on each variable. The combined uncertainty for Australia is much smaller than for India. This is in line with the tier-3 approach of Australia. Although India did report country-specific emission factors for underground mining based on measurements. It might therefore be possible to derive a reduced emission factor uncertainty, which is now by default taken as 100%, based on the individual measurements but these were not at our disposal. The combined uncertainty of ±80% is hence a conservative estimate and may possibly overestimate the real uncertainty. Nevertheless, with our estimate of 0.83 Tg yr-1 and adding 80% uncertainty shows our estimate is still significantly different from the EDGAR and GAINS estimates of 2.4 and 1.6 Tg yr-1, respectively.

## 5.2 Spatial distribution

The spatial distribution of coal mine methane emissions shows significant improvement for India and Australia using the bottom-up method and proxies described in section 4 (Figure 3 and 4).

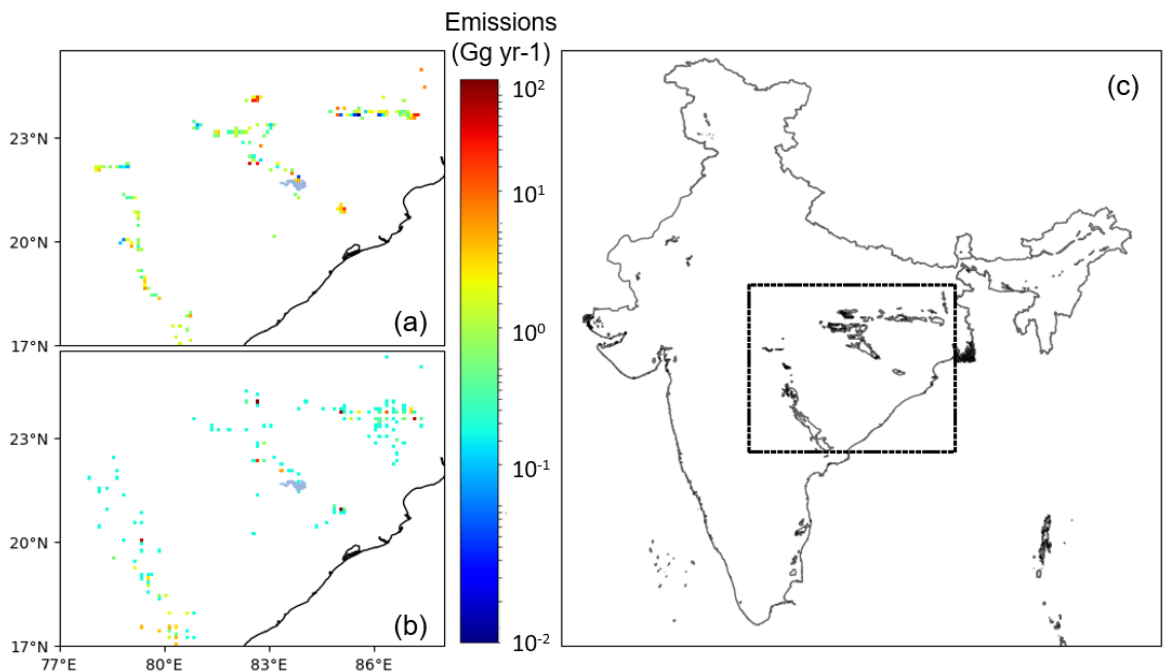

**Figure 3. Gridded bottom-up coal mine methane emissions from central India.**

Gridded coal mine methane emissions from (a) this study using raw coal production and emission factors. (b) EDGARv4.3.2 and, (c) Coal deposits in India and area of interest (dotted-box). All the emissions are gridded on a spatial resolution of 0.1° × 0.1°.

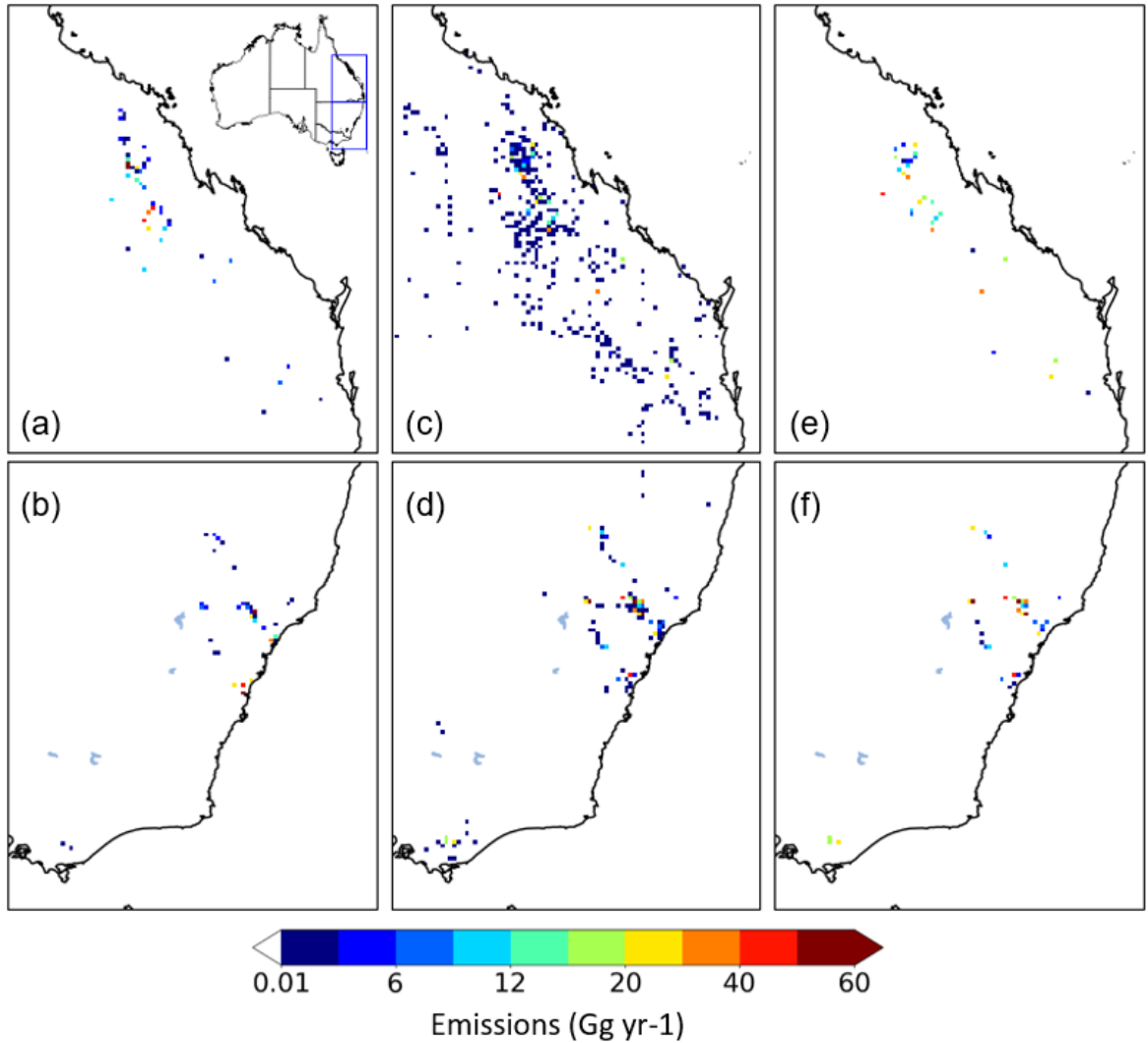

**Figure 4. Comparison of bottom-up coal mine methane emissions from Australia.**

Figure a-b shows the gridded (0.1° × 0.1°) coal mine methane emissions from Queensland, New South Wales and Victoria state from this study. Figure c-d shows the methane emissions from EDGARv4.3.2 for hard coal and brown coal. Figure e-f shows the methane emissions from EDGARv4.3.2 after removing the constant value (0.2481-0.2483) found in figure c-d.

Each grid encompasses the absolute location identified for surface and underground coal mines for both countries. These improvements can also be confirmed with the presence of coal deposits in India (Figure 3a and 3c), where the coal mines are located and emissions are distributed, rather than widely spaced spatial proxies used in EDGARv4.3.2 (Figure 3b). The comparison on a grid level shows spatial differences with global inventory of EDGARv4.3.2 for India and Australia (Figure 3 and 4). The grid-by-grid analyses explains two major differences, i) variation in high emitting grid cells i.e. coal production activity in each grid cell and ii) the difference in their spatial allocation i.e. location of mining operation. For India, in our inventory the highest methane grid cell emits 61 Gg yr-1 while the next top five grid cells emit between 33-44 Gg yr-1 (Figure 5) and account for 30% (highest six grid cells) of the total emission.

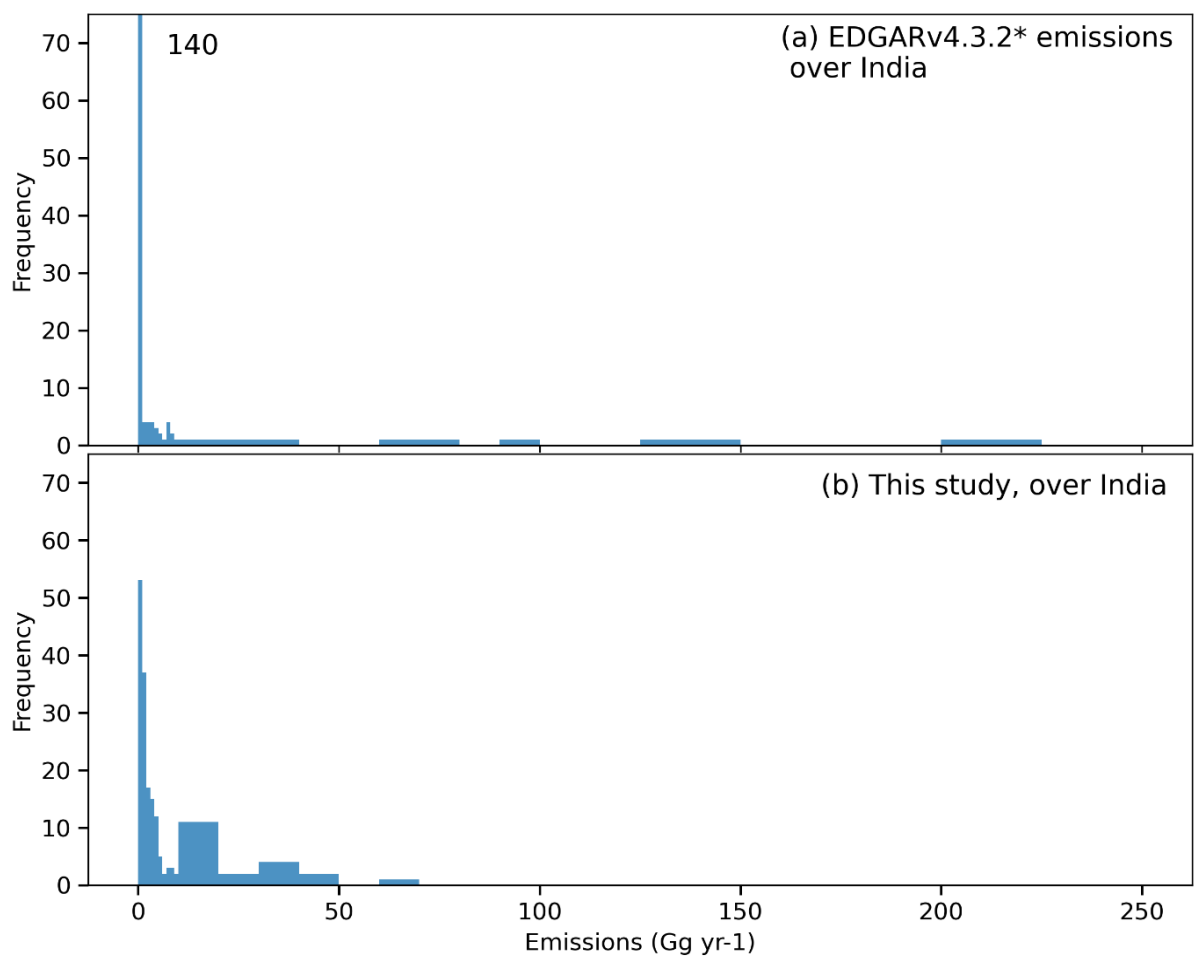

**Figure 5. Emissions frequency distribution over India.**

Figure shows the frequency plot of 0.1° × 0.1° gridded emissions from (a) EDGARv4.3.2* and (b) this study over India. * indicates the CMM emissions are scaled down, from 2310 Gg yr-1 to 825 Gg yr-1 from this study for comparison purpose. The emissions are binned in logbins starting 0-10 at an interval of 1, 10-100 at an interval of 10, and 100-250 at an interval of 25. Total number of gridded values in this study are 169 and in EDGAR are 172. The highest grid in this study emit 62 Gg yr-1 and the top six grid cell account for 31% of the coal mine methane emissions, while in EDGAR, the highest grid emits 605 Gg yr-1 and the top six grid accounts for 75% of the coal mine methane emissions.

While the highest methane emission grid cell in EDGARv4.3.2 emits 605 Gg yr-1 with next five grid cells emitting between 85-368 Gg yr-1. These highest six grid cells account for 75% of their total coal methane emissions. For comparison purpose, if we scale down the emissions from EDGARv4.3.2 (2333

Gg yr-1) to our estimate (825 Gg yr-1) and use the same spatial proxies of EDGARv4.3.2 (similar to Scarpelli et al., 2020), we arrive at the highest methane emission grid of 216 Gg yr-1 and the next top five grid emit 30-131 Gg yr-1 (Figure 5). Scarpelli et al. (2020) uses the spatial proxies from EDGARv4.3.2 and distributes 2016 CMM emissions as reported to UNFCCC. We therefore compare our gridded emissions with EDGAR only. Moreover, a difference in their spatial allocation of high emitting grid cells was also observed (Figure S5 and S6).

For Australia, the EDGAR global inventory estimates 1228 Gg yr-1 of methane from coal mine against 972 Gg yr-1 reported by NIR 2018 (NIR 2018, 2020). While the highest emitting grid cell from the new spatial distribution is 110 Gg yr-1 with highest six grids contributing to 45% emissions. Whereas the spatial distribution of EDGARv4.3.2 provides the highest methane emission grid of 74 Gg yr-1 with top 6 grids contributing to 28% (Figure 6).

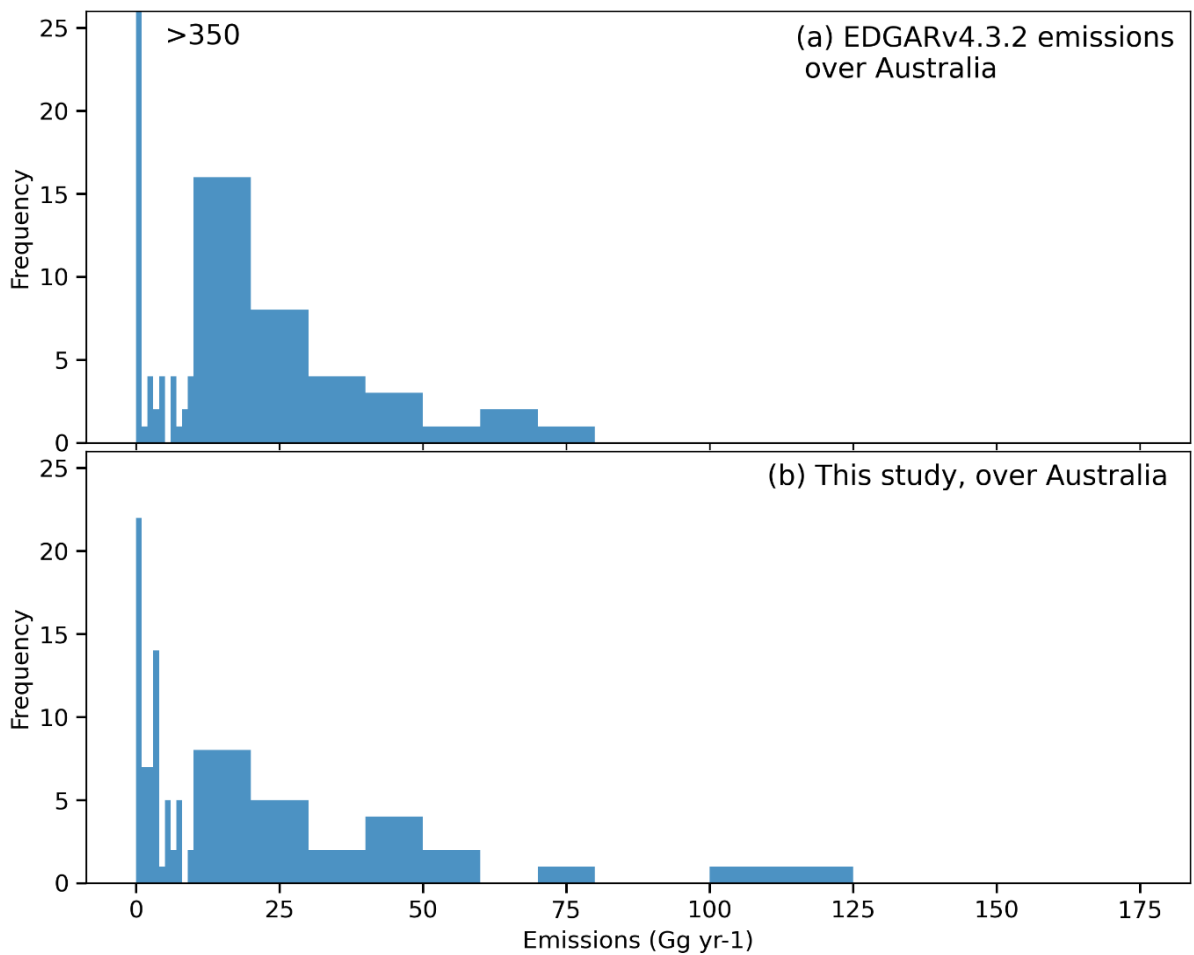

**Figure 6. Emissions frequency distribution over Australia.**

Figure shows the frequency plot of 0.1° × 0.1° gridded emissions from (a) EDGARv4.3.2 emissions and (b) this study over Australia. The emissions are binned in logbins as explained in figure 5. Total number of gridded values in EDGAR are 452 and in this study 88. In this study, the highest grid emits 110 Gg yr-1 and the top six grids account for 45% of the coal mine methane emissions. In EDGAR, the highest methane grid emits 74 Gg yr-1 and the top six grid accounts for 28% of the coal mine methane emissions.

A large number of low methane emission grid cells can be seen widely spread across all the coal producing states in EDGARv4.3.2 (Figure 4c-d). The methane emission in these grids range from 0.2481-0.2483 and account for 92 Gg yr-1 i.e. 9% of the national coal mine methane emissions. Although from Janssens-Maenhout et al. (2019) for EDGARv4.3.2 it is not clear what these low methane emission grid cells indicate. However, based on AUSGIN, (2020) and Australian mine atlas, the widely spread grids may refer to the location of mineral deposits (where mining has not yet started, but potential resources exists), historic mines and mines that have undergone care and maintenance, in addition to the operating coal mines. This work indeed only considers operating mines and distributes the emissions only to those locations. Figure 4e-f shows the spatial distribution after removing the low emission grids from Figure 4c-d, which probably represents the operating coal mines.

# 6   Conclusions

A fine resolution 0.1° × 0.1° bottom-up inventory was developed for the high coal producing countries of India and Australia for 2018. Together with China (Sheng et al., 2019) and US (Maasakkers et al., 2016) where the inventory already existed, these are the four largest coal producing countries. The emissions were estimated for each mine characterized by the type of mine and coal. Using the Indian coal production for each mine and the regionally measured tier-2 emission factors, emissions of 825 Gg yr-1 $CH_4$ was estimated for India for 2018. This estimate is well aligned with emissions estimate of 770 Gg yr-1 for 2012 and 788 Gg yr-1 for 2014 from previous Indian national reported emissions to UNFCCC. The emissions were spatially allocated to the identified locations of surface and underground coal mines, thereby improving the spatial distribution compared to the existing global inventory EDGARv4.3.2. The approach for Australia was slightly different. Australia is a so-called Annex-I country and annually reports its greenhouse gas emissions to the UNFCCC. These emissions are reported using tier-2/3 methodology as per the IPCC guidelines. In 2018, Australia reported 972 Gg yr-1 of methane emissions from coal mines grouped per coal-producing state in the categories of surface and underground mines. However, these emissions are not provided in gridded format. Australian CMM emissions are aggregated at the state level and not publicly available at the individual mine level. We focused our efforts for Australia on breaking down its reported emissions and improving the spatial allocation of emissions by compiling the location of surface and underground mines by type of coal. The state level emissions were further distributed to individual mines as per the amount of coal produced and the gas content profile of the underground mines. Like with India, the spatial distribution of Australian coal mine methane emission was significantly improved and differs substantially from the EDGARv4.3.2 distribution. In Australia, a shortcoming is that no clear information is available on the emission control strategies and efficiency by mine that are in place to reduce methane emissions publicly. These reductions are taken into account in the state-level total emission but because of lack of information at mine-level we assumed it is equally distributed over all mines within a state. The high resolution gridded emissions can be used as prior information in modelling and inversion studies to explain potential methane hotspots and trace back the underlying source. As a recommendation for future work we suggest validation and potential further improvement of magnitude by using recent satellite observations for example from TROPOMI on ESA's Sentinel-5 Precursor (Veefkind et al., 2012).

## Contributions

Contributed to conception and design: P.S, H.D.G.

Contributed to acquisition of data: P.S.

Contributed to analysis and interpretation of data: P.S.

Drafted and/or revised the article: P.S., S.P., S.H., J.M., H.D.G., I.A.

Approved the submitted version for publication: P.S., S.P., S.H., J.M., H.D.G., I.A.

## Acknowledgments
We thank Dr. Höglund-Isaksson, IIASA, Vienna for her comments on comparison of implied emission factors from Indian coal mines and Antoon Visschedijk, TNO, Utrecht for a productive discussion and expert comments on uncertainty calculations.

## Funding information
P.S. and S.P. are funded through the GALES project (#15597) by the Dutch Technology Foundation STW, which is part of the Netherlands Organization for Scientific Research (NWO).

## Competing interests
The authors declare no competing interests.

## Data accessibility statement
Final gridded emissions for India and Australia can be accessed here:
https://doi.org/10.5281/zenodo.6222441

**Supplemental Material**

A high-resolution gridded inventory of coal mine methane emissions for India and Australia

Pankaj Sadavarte[*], Sudhanshu Pandey, Joannes D. Maasakkers, Hugo Denier van der Gon, Sander Houweling, Ilse Aben

[1]SRON Netherlands Institute for Space Research, Utrecht, The Netherlands.

[2]Department of Climate, Air and Sustainability, TNO, Utrecht, The Netherlands.

[3]Department of Earth Sciences, Vrije Universiteit, Amsterdam, The Netherlands.

*Corresponding author: p.sadavarte@sron.nl

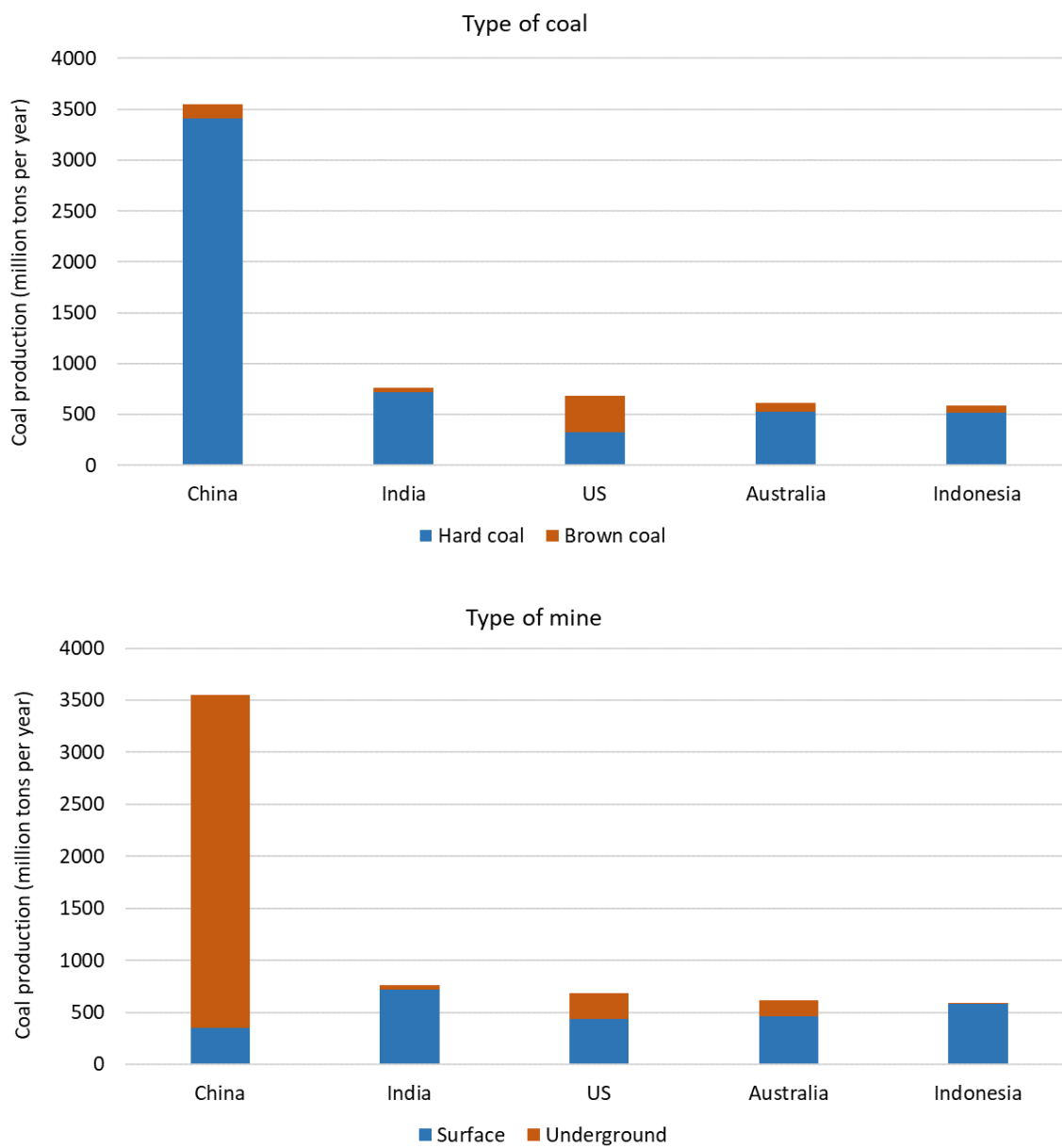

**Figure S1.** Top 5 coal producing countries in 2018. (a) Type of coal and (b) type of mine for top 5 coal producing countries in the year 2018. The annual coal production for China was compiled from Sheng et al., (2019), for India from provisional coal statistics (2018-19), MoC, 2019, for US from energy information administration (EIA) and for rest of the countries from https://yearbook.enerdata.net/coal-lignite/coal-production-data.html. The type of coal and type of mine categories were referred from global methane initiative country profiles report.

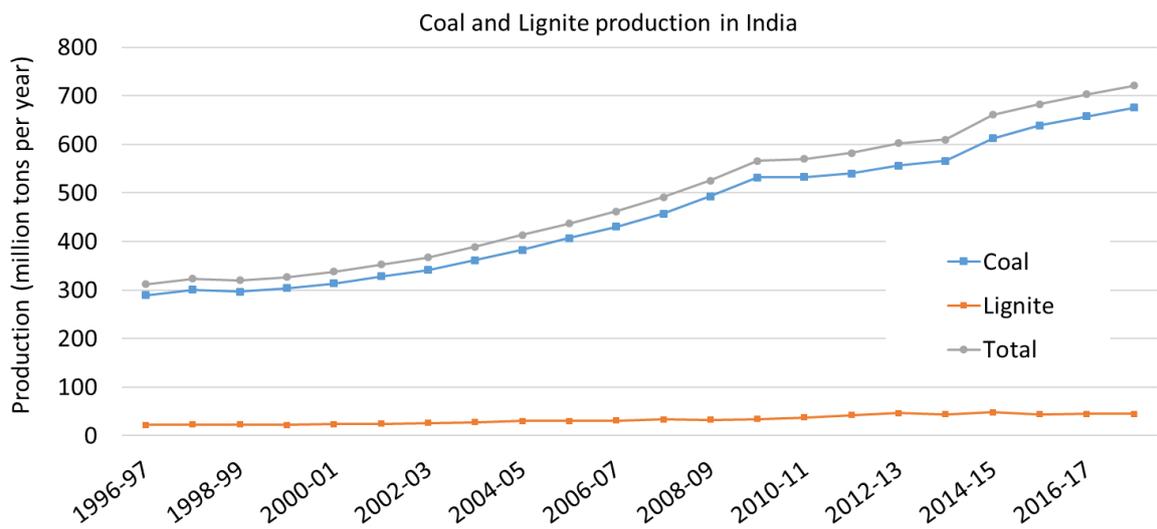

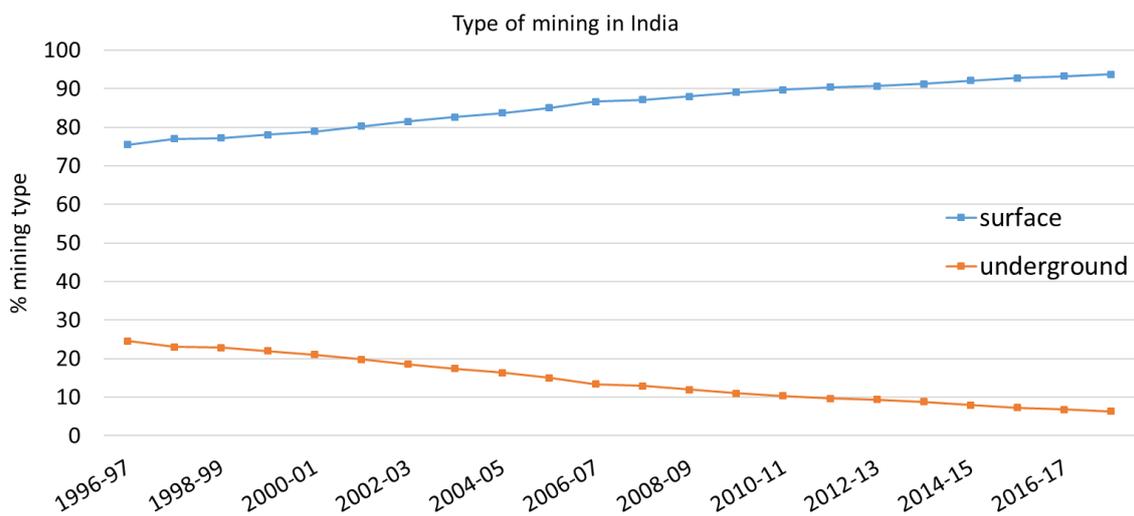

**Figure S2.** Trend in national coal production in India. Annual national coal production trend for India since 1996-97 in the categories of (a) types of coal – hard coal (bituminous/sub-bituminous) and brown coal (lignite) (b) type of mining – % national coal produced by surface and underground compiled from provisional coal statistics 2018-2019 report, MoC, 2019.

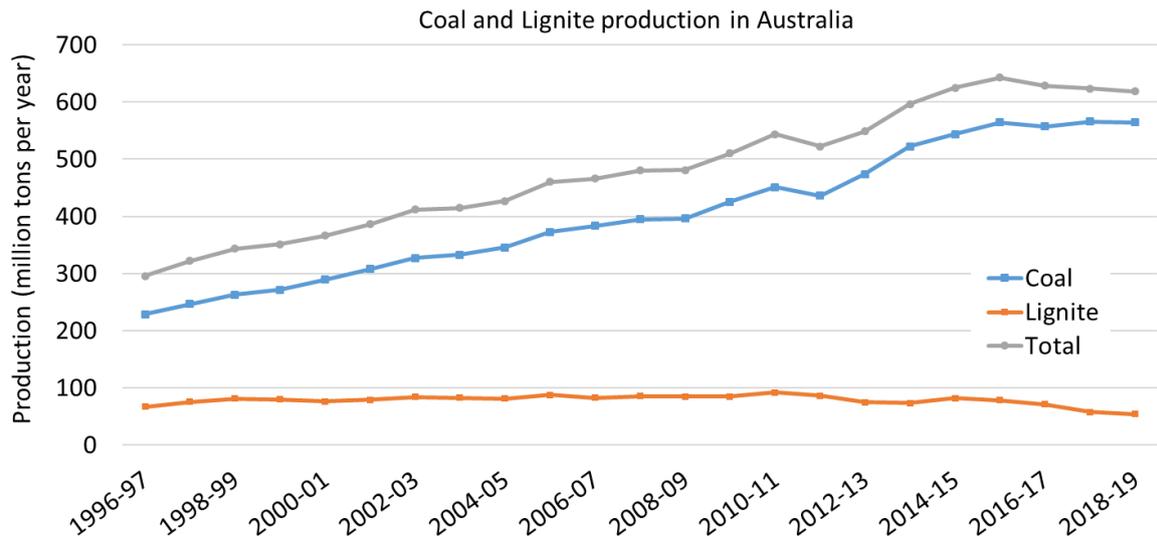

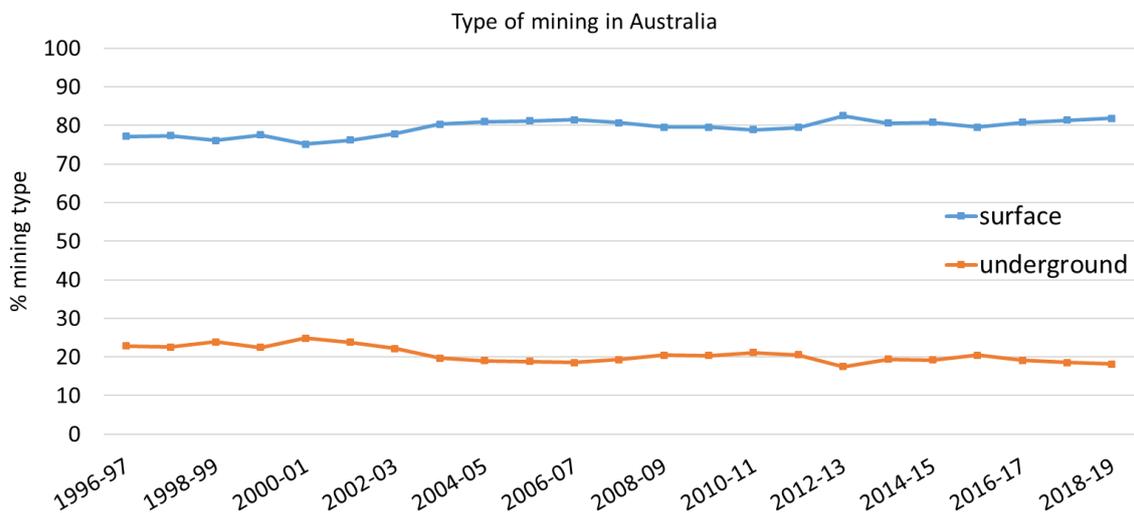

**Figure S3.** Trend in national coal production in Australia. Annual national raw coal production trend for Australia since 1996-97 in the categories of (a) types of coal – hard coal (bituminous/sub-bituminous) and brown coal (lignite) (b) type of mining – % national coal produced by surface and underground compiled from https://www.energy.gov.au/publications/australian-energy-update-2020, last accessed on 21st March 2021 and common report format (CRF) table 1.B.1 from national inventory report (NIR 2018, 2020).

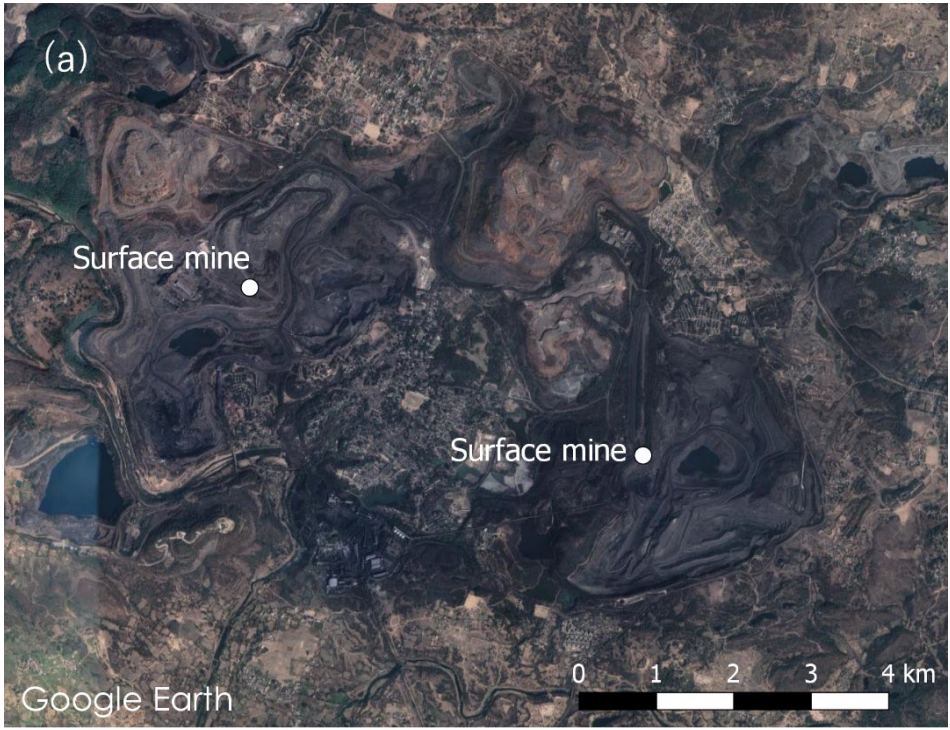
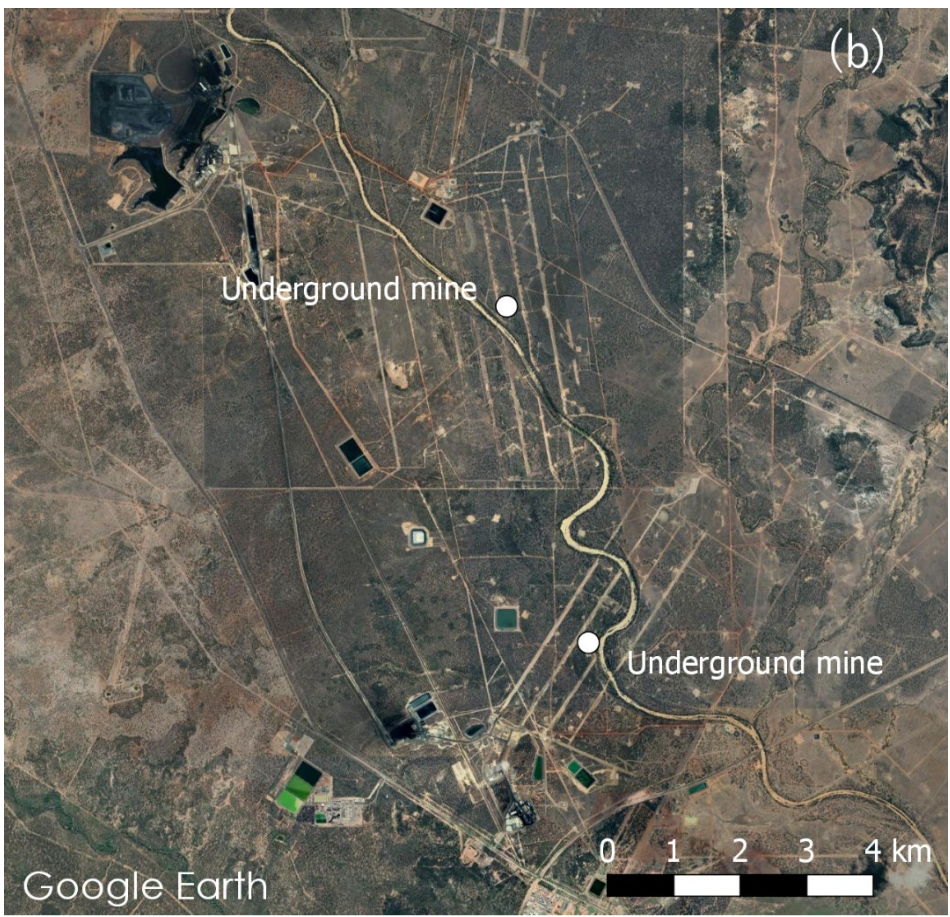

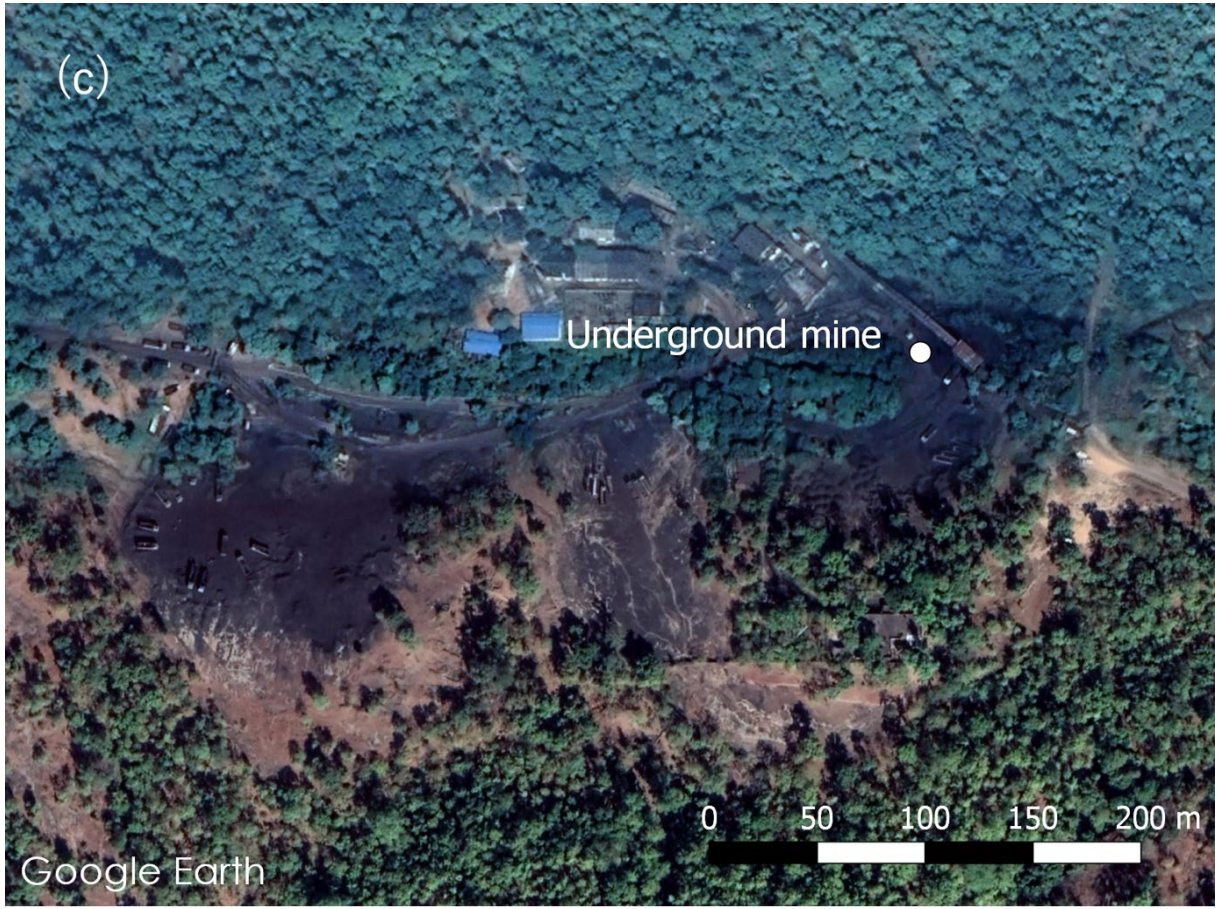

**Figure S4.** Surface and underground coal mines. (a) Shown are surface mines in India. Similar patterns are found for surface coal mines in Australia (b) Underground coal mine in Australia. Parallel lines are found near underground mines in Australia which shows the panels below the surface where the longwall mining is done (c) Underground coal mine in India, an incline conveyor infrastructure is usually seen that forms the entry and exit point for the people and coal.

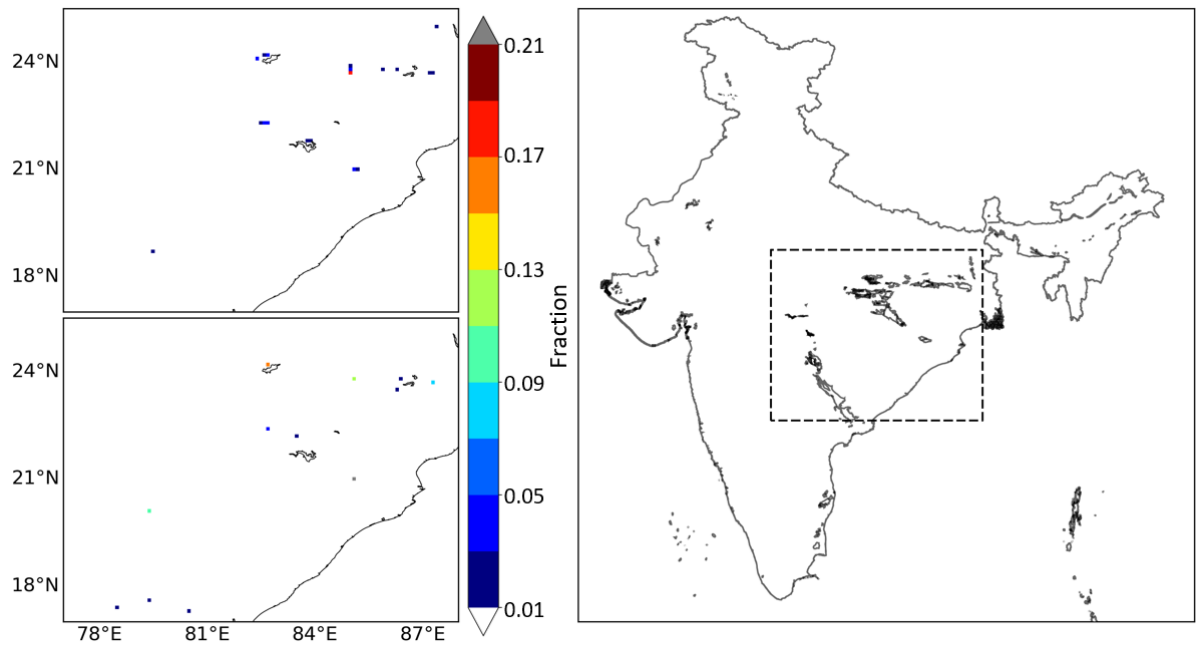

**Figure S5.** High methane emitting grids shown as fraction over India. Figures on the left top and bottom shows the methane emitting grids as fraction of the total coal mine methane emission from this study and EDGARv4.3.2. Large differences can be seen in the fractions and their spatial location. This study has identified location of surface and underground coal mines through various sources like Google Earth, annual reports and mining plans. Figure on the right shows the locations of coal deposits in India and the dotted box shows the region plotted for emission fraction comparison.

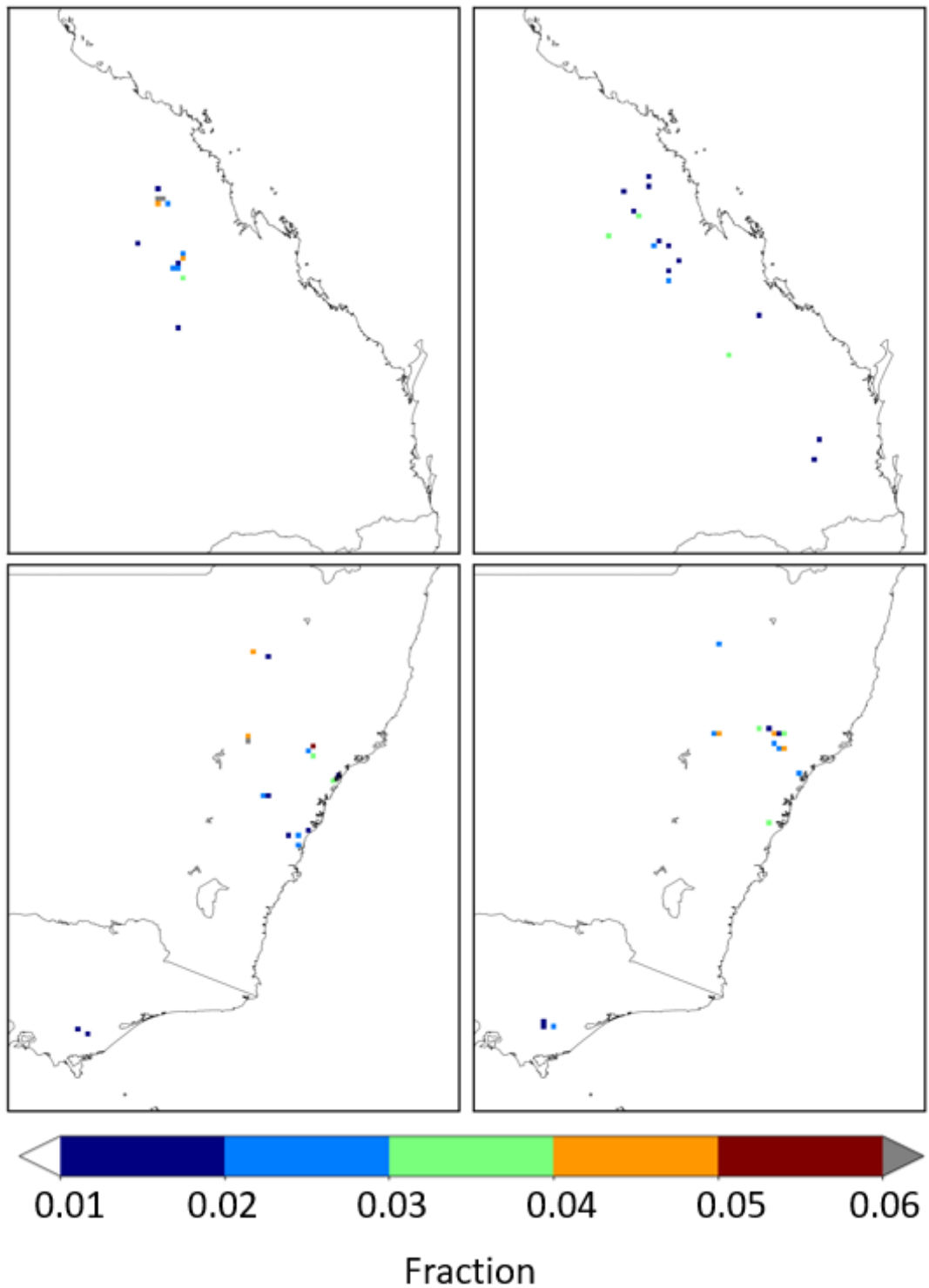

**Figure S6.** High methane emitting grids shown as fraction over Australia. Figures on the left top and bottom show the methane emitting grids from this study as fraction of the total coal mine methane emission over Queensland, New South Wales and Victoria from this study. Similarly, figures on right top and bottom are the emission fractions from EDGARv4.3.2. Large differences can be seen in the fractions and their spatial location. This study has identified locations of surface and underground coal mines through various sources like Google Earth, annual reports and mining plans. While EDGARv4.3.2 have re-distributed annual emissions as reported to UNFCCC using spatial proxies of EDGARv4.3.2.

**Table S1.** Coal mine methane emission estimates from various studies.

|  | Methane emissions (Gg yr-1) for 2012 | | | | |
| --- | --- | --- | --- | --- | --- |
|  | UNFCCC[c] | EDGAR v4.3.2[d] | GAINS[e] | CEDS[f] | Regional studies |
| **China** | 21015[a] | 21701 | 24570 | 37736 | 17200[g] |
| **US** | 2658 | 2527 | 2964 | 4594 | 2907[h] |
| **India** | 788[a] | 2333 | 1623 | 2569 | 770[i] |
| **Indonesia** | 80[b] | 2738 | 3400 | 1921 |  |
| **Australia** | 1186 | 1228 | 1188 | 1212 |  |

[a]Emissions from Second Biennial Update Report on Climate Change for base year 2014
[b]Emissions from Second Biennial Update Report on Climate Change for base year 2012
[c]Emissions compiled from National Inventory Report 2014 communicated to UNFCCC for base year 2012
[d]Emissions from EDGARv4.3.2 global bottom-up inventory of greenhouse gases for year 2012
[e]Annual emissions estimated using implied emission factor derived in Hoglund et al., 2012 and 2012 coal production data
[f]Hosley et al., 2018
[g]Sheng et al., 2019
[h]Maasakkers et al., 2016
[i]Singh et al., 2015

**Table S2.** Implied emission factor for Indian and Australian coal mines.

|  | Year | Underground | Surface | National |
|---|---|---|---|---|
| **IPCC default**[a] |  |  |  |  |
| < 200m |  | 7.38 | 0.2 |  |
| 200-400m |  | 13.87 | 0.88 |  |
| > 400m |  | 19.62 | 1.49 |  |
|  |  |  |  |  |
| **Kholod et al., 2020**[b] |  |  |  |  |
| < 200m |  | 10.08 | 2.03-3.38 |  |
| 200-400m |  | 12.79 |  |  |
| > 400m |  | 14.62 |  |  |
| Brown coal |  |  | 0.52 |  |
|  |  |  |  |  |
| **India** |  |  |  |  |
| This study, Black coal | 2018 | 5.05 | 0.87 | 1.04 |
| Brown coal | 2018 |  | 0.22 |  |
| EDGARv4.3.2[c] | 2012 |  |  | 3.9 |
| EDGARv5.0[d] | 2015 |  |  | 3.9 |
| GAINS[e] | 2012 |  |  | 3 |
| CEDS[f] | 2012 |  |  | 4.3 |
|  |  |  |  |  |
| **Australia** |  |  |  |  |
| NIR[g] | 2018 | 6.31 | 0.52 | 1.6 |

All units in kton $CH_4$ per million ton raw coal.
[a]2006a IPCC guidelines for national greenhouse gas inventories, Vol. 2. Energy, Chapter 4, Fugitive emissions.
[b]Kholod et al., 2020.
[c]EDGARv4.3.2 CH4 emissions for 2012 (Janssens-Maenhout, et al., 2019) and raw coal for corresponding year from provisional coal statistics 2018-2019.
[d]EDGARv5.0 CH4 emissions for 2015 and raw coal for corresponding year from provisional coal statistics 2018-2019.
[e]Höglund-Isaksson, 2012 and raw coal for corresponding year from provisional coal statistics 2018-2019.
[f]Hosley et al., 2018 and raw coal for corresponding year from provisional coal statistics 2018-2019.

**Table S3.** Uncertainty assessment for India and Australia coal mines.

| Coal mines | CH4 (Gg yr-1) | Uncertainty (%) in coal production | Uncertainty in emission factors | Propagated uncertainty | Fraction of total methane emission | Total uncertainty |
|---|---|---|---|---|---|---|
| **Australia** | | | | | | |
| Underground | 710.4 | 2[a] | 10[a] | 10.20[a] | 7.45[e] | |
| Surface | 262.6 | 6.30[a] | 30.89[a] | 31.53[a] | 8.36[e] | 11.20[e] |
| Others | | 5.00[a] | 50[a] | 50.25[a] | 0.20[e] | |
| **India** | | | | | | |
| Underground | 606.40 | 10.0[b] | 100.0[d] | 100.50 | 75.97[e] | 79.83[e] |
| Surface | 195.81 | 10.0[b] | 100.0[c] | 100.50 | 24.53[e] | |

[a]National inventory report, 2018, Volume-III, Annexes, page 137.
[b]Solazzo et al., 2021, Table 1.
[c]Third Biennial Update Report to The United Nations Framework Convention on Climate Change, Table 2.34.
[d]Assumed similar to surface mining in India.
[e]Calculated using IPCC Quantifying Uncertainties in Practice (Volume 1 Chapter 3 IPCC 2006b).